\documentclass[aps,twocolumn]{revtex4}
\usepackage{graphicx}
\usepackage{epstopdf}
\newcommand{\vect}[1]{\mbox{\boldmath $#1$}}
\newcommand{\D}{{\rm d}}
\newcommand{\E}{{\rm e}}

\begin{document}

\title{Perturbation Theory for Traveling Droplets}
\author{L.M. Pismen \\
{\it Department of Chemical Engineering and 
Minerva Center for Nonlinear Physics of Complex Systems,\\
Technion -- Israel Institute of Technology, 32000 Haifa, Israel}} 
\date{\today}

\begin{abstract}
Motion of chemically driven droplets is analyzed by applying a solvability condition of perturbed hydrodynamic equations affected by the adsorbate concentration. Conditions for traveling bifurcation analogous to a similar transition in activator-inhibitor systems are obtained. It is shown that interaction of droplets leads to either scattering of mobile droplets or formation of regular patterns, respectively, at low or high adsorbate diffusivity. The same method is applied to droplets running on growing terrace edges during surface freezing. 
\end{abstract}

\pacs{68.15.+e, 47.20.Ky, 68.43.-h}
\maketitle

% %%%%%%%%%%%%%%%%%%%%%%%%%%%%%%%%%%%%%%%%%%%%%%%%%
\section{Introduction }
% %%%%%%%%%%%%%%%%%%%%%%%%%%%%%%%%%%%%%%%%%%%%%%%%%
%
Spontaneous motion of droplets on solid substrate driven by chemical reactions that influence wetting properties has been observed in a number of experiments \cite{e1,e2,e3,e4}. More recently, a reversible setup allowing for restoration of substrate properties and thereby making permanent motion possible has been realized experimentally \cite{j05l,j05e}. Spontaneous motion has been observed also in processes of surface freezing and melting \cite{r05}. The cause of chemically driven motion is deposition of a poorly wettable material on the substrate or, alternatively, dissolution of an adsorbed wettable layer beneath the droplet. As a result, the droplet tends to migrate to the area with more favorable wetting properties outside its footprint. The direction of motion, originally chosen at random, is sustained because the area left behind is either permanently left nonwettable or takes time being restored. Chemically driven motion is of interest both in microfluidics applications and as a possible mechanism for the formation of surface patterns.   

The theoretical model of chemically driven droplet motion presented recently \cite{th04,th05} combined hydrodynamic equations in lubrication approximation with a linear reaction-diffusion equation for adsorbed species. The model included a precursor layer that served to resolve the contact line singularity. The model equations were solved numerically, mapping the regimes of motion and its dependence on various parameters of the problem. The results showed, in particular, that droplets become immobile when diffusion and substrate modification are sufficiently fast. The computations were, however, restricted to moderate ratios of the bulk droplet size to the precursor layer thickness,  since numerical routines are all but impossible to implement for realistic very large ratios. 

In this communication, I present an analytical solution of the problem combining an integral solution of the reaction-diffusion equation for a steadily propagating droplet \cite{p01} with velocity computation using the solvability condition of perturbed hydrodynamic equations. The hydrodynamic problem is considered in Section~\ref{S0}. The approach, based on the lubrication approximation, is the same as in the theory of creeping motion of droplets under the action of externally imposed gradients. The perturbation approach to the problem of motion driven by surface inhomogeneities was pioneered by Greenspan \cite{g78} who had, however, to rely on a phenomenological relation for the motion of the contact line. Brochard \cite{b89} derived droplet translation velocity using integral balances between various driving forces and viscous dissipation. The latter poses a special problem in view of the notorious contact line singularity, which can be resolved either by introducing a slip length \cite{h83,eg05f} or by allowing for an ultrathin precursor layer \cite{pp00}.
The latter approach has been applied to derive integral conditions for a droplet  driven either by an external field (gravity) or changes in the precursor thickness due to droplet interactions \cite{pp04}. The result, applied later also to 2D droplets driven by a difference of advancing and receding contact angles, turned out to give even better approximation than more elaborate theory using a precise solution of lubrication equations to compute the shape of the bulk droplet \cite{tp06}. 

The integral condition for droplet motion will be derived in Section~\ref{S1} in a more formal way as a solvability condition of perturbation equations, using an eigenfunction of the adjoint problem introduced earlier for a 2D problem \cite{gl03}. The contact line singularity will be resolved, however, in a more traditional way through introducing a slip length, and the solvability condition will be re-derived for this model. Both approaches to eliminating the singularity are indistinguishable on a macroscopic level, leading to a model-dependent logarithmic factor \cite{eg05p}, and therefore the result does not need to rest on the existence of a macroscopic precursor or depend on a precise way the hydrodynamic equations have to be modified in the immediate vicinity of the substrate.         

In Sections~\ref{S3} and \ref{S4}, I shall concentrate on the reversible setup of Refs.~\cite{j05l,j05e}. The basic approach is described in Section~\ref{S0}, followed by the velocity computation for a single steadily propagating droplet both in fast (Section~\ref{S31}) and slow (Section~\ref{S34}) diffusion limits and in a general case (Section~\ref{S36}). The principal result obtained in the fast diffusion (slow velocity) limit is the existence of a supercritical traveling bifurcation analogous to a similar transition in activator-inhibitor systems \cite{km94,book}. I further investigate droplet interactions on either side of this transition, resulting in relaxation to a regular stationary pattern sustained by long-range repulsion (Section~\ref{S42}) or scattering of mobile droplets (Section~\ref{S41}). The problem of motion driven by surface freezing or melting is briefly considered in Section~\ref{S6}. 
%         

% %%%%%%%%%%%%%%%%%%%%%%%%%%%%%%%%%%%%%%%%%%%%%%%%%
\section{Solution Method  \label{S0}}
% %%%%%%%%%%%%%%%%%%%%%%%%%%%%%%%%%%%%%%%%%%%%%%%%%
%
% %%%%%%%%%%%%%%%%%%%%%%%%%%%%%%%%%%%%%%%%%%%%%%%%%
\subsection{Lubrication Equations  \label{S01}}
% %%%%%%%%%%%%%%%%%%%%%%%%%%%%%%%%%%%%%%%%%%%%%%%%%
%

The droplet shape is described in the lubrication approximation by the thin film equation
\begin{equation}
h_t = -\frac{\gamma}{\eta}\, \nabla\cdot \left[ q(h)\, 
\nabla\, \nabla^2 h  \right]  .
\label{film}
\end{equation}
Here gravity and other external forces are neglected, $\gamma$ is the surface tension of the droplet interface, and $\nabla$ is the 2D gradient operator in the plane of the substrate. The simplest suitable expression for the effective mobility function $q(h)$, obtained assuming the viscosity $\eta$ of the droplet to be much larger than that of the surrounding fluid and applying the Navier slip boundary condition, is 
\begin{equation}
 q(h) = \frac{h^2}{3} \,\left( h+ 3\lambda \right) ,
\label{slip}
\end{equation}
where $\lambda$ is the slip length. 

The boundary condition on the droplet contour $\Gamma$, i.e.\ the contact line, is $\vect{n} \cdot\nabla h = - \theta$, where $\vect{n}$ is the outer normal to $\Gamma$ and $\theta$ is the contact angle. The droplet is stationary when the  equilibrium contact angle does not depend on position explicitly. An asymmetry of the contact angle caused by the substrate modification sets the droplet into motion. We shall assume that the asymmetry is weak, so that the motion is slow and the change of the droplet shape can be viewed as a small correction. 

Taking the direction of motion as the $x$ axis, we rewrite Eq.~(\ref{film})
in the comoving frame in the dimensionless form
\begin{equation}
\delta\,\widehat{\vect{x}} \cdot\nabla h = \nabla\cdot \left[ q(h)\, \nabla\, \nabla^2 h  \right]  .
\label{film1}
\end{equation}
where $\delta = U\eta/\gamma$ is the capillary number based on the droplet velocity $U$, as yet unknown, and $\widehat{\vect{x}}$ is the unit vector in the direction of motion. The length scale in this equation remains arbitrary. 

Assuming $\delta \ll 1$, we expand
\begin{equation}
 h = h_0 + \delta h_1 + \ldots ,  \quad
  \theta = -\vect{n} \cdot\nabla \left(h_0 +\delta h_1 + \ldots \right)_\Gamma.
\label{del}
\end{equation}

The zero order function is the stationary solution which verifies the Laplace equation $\nabla^2 h_0=0$. The solution with a constant contact angle $\theta_0$ is just a paraboloidal cap  
\begin{equation}
  h_0(r)  = \frac{R \theta_0}{2} \left[1 - \left(\frac{r}{R} \right)^2\right],
\label{film0}
\end{equation}
where $r$ is the radial coordinate and $R$ is the droplet radius. The perturbed shape of the moving droplet should be obtained from the first-order equation obtained by expanding Eq.~(\ref{film1}) in $\delta$. It turns out, however, that a relation between the velocity and the contact angle distortion can be obtained without actually solving this equation; it is sufficient to compute its \emph{solvability condition}.

% %%%%%%%%%%%%%%%%%%%%%%%%%%%%%%%%%%%%%%%%%%%%%%%%%
\subsection{Translational Solvability Condition \label{S1}}
% %%%%%%%%%%%%%%%%%%%%%%%%%%%%%%%%%%%%%%%%%%%%%%%%%
%
The first-order equation has a general form
\begin{equation}
{\cal L}h_1 + \Psi(\vect{x}) = 0,  
\label{h1eq} \end{equation}
which contains the linear operator
\begin{equation}
{\cal L}h_1 \equiv  -\nabla \cdot \left[ q(h_0)  \nabla \nabla^2 h_1\right]  
\label{h1op} \end{equation}
and the inhomogeneity
\begin{equation}
\Psi(\vect{x}) = \widehat{\vect{x}}\cdot \nabla h_0.
\label{inhom}  \end{equation}
The operator ${\cal L}$ is not self-adjoint. The adjoint equation defining the translational Goldstone mode $\varphi$ is
\begin{equation}
{\cal L}^\dag \varphi =  - \nabla^2 \left[  \nabla \cdot   
q(h_0)\, \nabla \varphi \right] =0 .  
\label{h1opadj} \end{equation}
This equation is verified by the eigenfunction 
\begin{equation}
 \varphi =  \int \frac{h_0}{q(h_0)}\, \D x .%, \qquad  
%\nabla \varphi =  \widehat{\vect{x}} \frac{h_0}{q(h_0)} ,   
\label{eigen} \end{equation}
The integration can be carried out along an arbitrary axis $x$, which can be chosen to coincide with the direction of motion. Integrating along the two Cartesian axes gives two Goldstone modes corresponding to two translational degrees of freedom in the plane.

The solvability condition of Eq.~(\ref{h1eq}) defining the translation speed is obtained by multiplying it by $\varphi$ and integrating over the droplet footprint $\cal R$ bounded by a contour $\Gamma$. Since the solvability condition is evaluated in a finite region, it includes both the area and contour integrals. The area integral stemming from the inhomogeneity is evaluated using integration by parts: 
\begin{equation}
- {\cal J}= \int_{\cal R} \varphi(\vect{x}) \,
  \widehat{\vect{x}}\cdot \nabla h_0  \,\D \vect{x} 
%\nonumber \\ &=&
= - \int_{\cal R} \frac{h_0^2}{q(h_0)} \, \D \vect{x}.
 \label{fric} \end{equation}
The integral ${\cal J}$ is interpreted as the \emph{friction factor}. The divergence of this integral at $\lambda=0$ is the reason for introducing the slip length in Eq.~(\ref{slip}). Since, however, this length is very small, being measured on molecular scale, the integral can be evaluated by separating it into two parts. Near the contact line, i.e. in a ring $R \leq r \leq l$ where $\lambda \ll l \ll R$, the integration can be carried out using the linearized profile $h =\theta (R -r)$. This yields, asymptotically at $l \gg \lambda$,  
\begin{equation}
 {\cal J}_1= 6\pi R \int_{R-l}^R  
 [\theta_0(R-r) + 3\lambda]^{-1} \, \D r
 \asymp \frac{6\pi R}{\theta_0} \ln \frac{\theta_0 l}{3\lambda}.
 \label{fric1} \end{equation}
In the bulk region $r \leq R-l$, $\lambda$ can be neglected, and the integration  yields, asymptotically at $l \ll R$, 
\begin{equation}
 {\cal J}_2= \frac{6\pi}{ R\theta_0} \int_0^{R-l}  
 \left[1 - \left(\frac{r}{R} \right)^2\right]^{-1} \,r\, \D r
 \asymp \frac{6\pi R}{\theta_0} \ln \frac{R}{2l}.
 \label{fric2} \end{equation}
When both integrals add up, the auxiliary length $l$ falls out, resulting in an expression containing the logarithm of the ratio of the macroscopic and microscopic scales:
\begin{equation}
 {\cal J}=  \frac{6\pi R}{\theta_0} \ln \frac{\theta_0 R}{6\lambda}.
 \label{fricj} \end{equation}

An additional contour integral dependent on the unknown first-order function $h_1$ is contributed by the operator $\cal L$ when it is multiplied by $\varphi$ and integrated by parts:
%\samepage{ 
%\begin{widetext}
\begin{eqnarray}
{\cal I}_\Gamma &=& \!
- \oint_\Gamma \varphi(s)\, k(h_0)\,
\vect{n}\cdot \nabla \nabla^2 h_1 \D s 
%\nonumber \\ &+&
+ \oint_\Gamma  (\vect{n} \cdot \widehat{\vect{x}}) \, h_0
 \nabla^2 h_1 \D s \nonumber \\% \label{solconcont}\\ &-&
 &-& \! \oint_\Gamma (\widehat{\vect{x}} \cdot  \nabla h_0 )\, \vect{n} \cdot \nabla h_1  \,\D s 
+ \oint_\Gamma  (\vect{n}\cdot \widehat{\vect{x}})\,  \nabla^2 h_0 \, h_1 \D s. %\nonumber
\label{solconcont} \end{eqnarray} 
%\end{widetext}
%
The first integral vanishes, since at $h \to 0$ $q(h) \propto h^2$, while $\varphi(h)$ is only logarithmically divergent. The second integral vanishes at $h_0=0$ as well, and so does the last integral, since $h_0$ is harmonic. The remaining integral expresses the driving force due to the variable part of the contact angle $\widetilde \theta = \theta -\theta_0 = -\delta^{-1}\vect{n} \cdot \nabla h_1 $. Using the identity $\widehat{\vect{x}} \cdot  \nabla h_0 =-\theta_0 \cos \phi$, where $\phi$ is the polar angle counted from the direction of motion, the force is evaluated as 
\begin{equation}
{\cal I}_\Gamma =\frac{ a \theta_0}{\delta}
 \int_{-\pi}^{\pi} \cos \phi \, \widetilde \theta(\phi) \D \phi 
 \equiv \frac{\cal F}{\delta}  .
 \label{fricI} \end{equation}
Thus, the solvability condition defining the droplet velocity reads
\begin{equation}
\delta= \frac{\cal J}{\cal F} = \frac{\theta_0^2}{6\pi} 
 \ln^{-1} \frac{\theta_0 R}{6\lambda} \int_{-\pi}^{\pi} \cos \phi \, \widetilde \theta(\phi) \D \phi.
 \label{fricd} \end{equation}
%
% %%%%%%%%%%%%%%%%%%%%%%%%%%%%%%%%%%%%%%%%%%%%%%%%%
\section{Chemical Self-Propulsion  \label{S3}}

% %%%%%%%%%%%%%%%%%%%%%%%%%%%%%%%%%%%%%%%%%%%%%%%%%
\subsection{Surfactant distribution \label{S2}}
% %%%%%%%%%%%%%%%%%%%%%%%%%%%%%%%%%%%%%%%%%%%%%%%%%

\begin{figure} [t]
\begin{center}
\includegraphics[width=8.5cm]{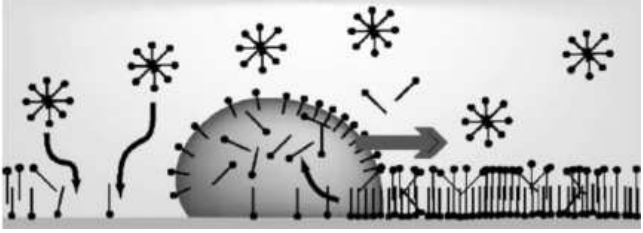} 
\end{center}
\caption{Schematic diagram of droplet motion due to modification of surface wettability \cite{j05l,j05e}. The adsorbed surfactant is represented as a hydrophobic bar with a hydrophilic head. \label{fj}}
\end{figure}

%
Variation of the contact angle is caused by substrate modification, e.g. dissolution of the surfactant adsorbed on the substrate in experiments of Sumino et al. \cite{j05l,j05e} (Fig.~\ref{fj}). We write the the adsorption-diffusion equation for the surfactant coverage on the substrate in the dimensionless form
\begin{equation}
c_t = \nabla^2 c - c +  H(\vect{x}) . 
\label{diff} \end{equation}
Here $H(\vect{x})$ equals to 1 outside and 0 inside the droplet footprint. The surfactant coverage $c$ is scaled by the coverage in equilibrium with the surfactant concentration in the continuous phase, time by the inverse adsorption/desorption rate constant $k$ (which are assumed for simplicity to be equal), and length, by $\sqrt{D/k}$, where $D$ is the surfactant diffusivity on the substrate. 

The equation transformed to the frame moving with a dimensionless velocity $v=U/\sqrt{D k}$ along the $x$ axis is
\begin{equation}
vc_x + \nabla^2 c - c +  H(\vect{x}) =0. 
\label{eqmot} \end{equation}
The solution of this equation can be expressed with the help of an appropriate Green's function through an integral over the droplet footprint area and subsequently transformed into a contour integral with the help of the Gauss theorem \cite{p01}.  For a circular contour with a dimensionless radius (Thiele modulus) $a=R \sqrt{k/D}$, the concentration on its boundary, which determines the propagation speed, is computed in this way as  
\begin{widetext}
\begin{eqnarray}
 c(\phi) = 1 &-& \frac{a}{2\pi}  \int_{-\pi}^{\pi} 
   \E^{-\frac{1}{2} v a(\cos \phi- \cos \xi)} %\; \times \cr &&
 \left[ \frac{ v}{2} \cos \xi \,
 K_0\left(2a \sqrt{1+  \frac{v^2}{4}}
\sin \frac{|\phi-\xi|}{2}  \right) \right. \nonumber \\ 
& + &\sqrt{1+  \frac{v^2}{4}}\, \sin  \frac{|\phi-\xi|}{2}  
\left. K_1\left(2a \sqrt{1+  \frac{v^2}{4}}
\sin \frac{|\phi-\xi|}{2}  \right)  \right] \D \xi,
\label{woint1}  \end{eqnarray}
\end{widetext}
where $K_n$ (and $I_n$ below) are modified Bessel functions. 

Assuming the contact angle to be a linear function of the surfactant coverage, $\widetilde \theta = - \beta c$, the propagation velocity is computed by solving the equation obtained by combining Eqs.~(\ref{fricd}) and (\ref{woint1}):
\begin{equation}
v =  \frac{ M}{\pi}  \int_0^{\pi}\widetilde{c}(\phi;v)  \cos \phi \,\D \phi,
 \label{fricv} \end{equation}
where $\widetilde{c}=1-c$ and all relevant parameters, except the droplet radius $a$ in Eq.~(\ref{woint1}), are lumped into a single dimensionless combination, which can be called the mobility parameter:
\begin{equation}
M = \frac{\theta_0^2  \sigma \beta}{6 \eta \sqrt{D k}} 
 \ln^{-1} \frac{\theta_0 R}{6\lambda}.
 \label{fricm} \end{equation}
The parameter $M$ retains a weak logarithmic dependence on the droplet radius. %, w additional dependence on $a$ comes from Eq.~(\ref{woint1}).

%\subsection{Fast Diffusion Limit \label{S31}}
% %%%%%%%%%%%%%%%%%%%%%%%%%%%%%%%%%%%%%%%%%%%%%%%%%
\subsection{Traveling Bifurcation \label{S31}}
Simplified expressions can be obtained in the limiting cases $v \ll 1$ and  $v \gg 1$. The parameter $\delta$ can be still small also in latter case, since the ratio of the characteristic ``chemical'' velocity $\sqrt{D k}$ to the characteristic velocity $\sigma/\eta$ which determines the influence of viscous stresses on the droplet shape is typically very small.

\begin{figure} [b]
\begin{center}
\includegraphics[width=8.5cm]{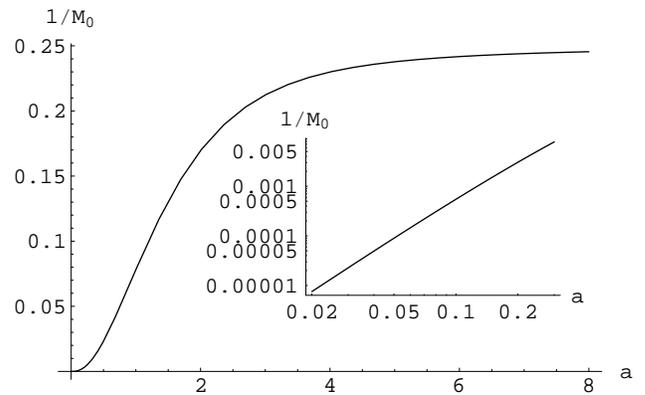} 
\end{center}
\caption{The critical value $M_0^{-1}$ as a function of the dimensionless droplet radius $a$. Inset: blow-up near the origin on logarithmic scale. The droplets are mobile below this curve. \label{fma}}
\end{figure}

The limit $v \ll 1$ (i.e.\ fast diffusion) is analogous to the fast inhibitor limit in reaction-diffusion systems, which is conducive to formation of stationary patterns \cite{book}. Equation (\ref{woint1}) is expanded in this limit to the first order in $v$ as
%
%{\nopagebreak 
\begin{widetext}
\begin{eqnarray}
\widetilde{c} (\phi)=  \frac{a}{\pi} \int_{0}^{\pi}\left\{\sin \frac{\zeta}{2}\, K_1\left(2a\sin \frac{\zeta}{2}\right) 
%\nonumber \\ 
  + v \cos \phi \left[\frac{1}{2}\cos \zeta\, K_0\left(2a\sin \frac{\zeta}{2}  \right) + a \sin^3 \frac{\zeta}{2}\, K_1\left(2a\sin \frac{\zeta}{2}  \right)  \right] + O(v^2) \right\} \D \zeta,
\label{woint0}  \end{eqnarray}
\end{widetext}
where $\zeta=\xi-\phi$. In the leading $O(1)$ order, the surfactant distribution is circularly symmetric. The first-order dipole term in Eq.~(\ref{woint0}) is the only one contributing to the integral in Eq.~(\ref{fricv}) (another term vanishing upon integration is omitted). The angular integrals are evaluated  using the identities
\begin{eqnarray*}
\Phi_k (a) &=& \!\int_{0}^{\pi} \sin^{2k} \frac{\phi}{2}\,
  K_0 \!\left( 2 a\sin \frac{\phi}{2} \right) \D\phi = \!
-\frac{1}{2a}\, \frac{\D(a\Psi_{k-1})}{\D a},  \\
\Psi_k(a) &=& \int_{0}^{\pi} \sin^{2k+1} \frac{\phi}{2}\,
  K_1\left( 2 a\sin \frac{\phi}{2}\right) \D \phi = 
- \frac{1}{2}\frac{\D \Phi_k}{\D a}, 
\end{eqnarray*}
starting from $\Phi_0(a)= \pi I_0(a) K_0(a)$. Plugging the resulting expressions in Eq.~(\ref{fricv}) yields the condition for the onset of motion
 \begin{equation}
 M_0^{-1}= \frac{a^2}{2} \left[I_1(a)K_2(a) - I_0(a)K_1(a) \right].                
   \label{onset} \end{equation}
The critical value $M_0^{-1}$ as a function of $a$ is plotted in Fig.~\ref{fma}.
Since the radial dependence in Eq.~(\ref{onset}) saturates when the droplet radius far exceeds the diffusional range, so that $M_0 \to 4$ at $a \to \infty$,
no droplet can move below this limiting value. 

\begin{figure} [t]
\begin{center}
\includegraphics[width=8.5cm]{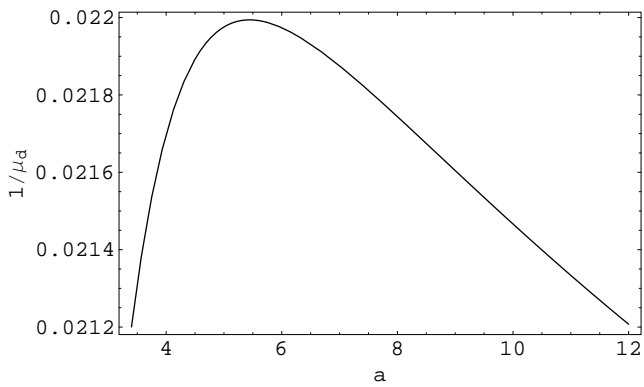} 
\end{center}
\caption{The critical value of $\mu_d^{-1}$ as a function of the dimensionless droplet radius $a$ for $\lambda_d =10^{-4}$. \label{fmua}}
\end{figure}

Because of the logarithmic dependence of $M$ on the droplet radius, the size dependence of the traveling threshold in Eq.~(\ref{onset}) still remains implicit. An explicit dependence can be extracted after rewriting Eq.~(\ref{fricm}) as 
\begin{equation}
M = \mu_d \ln^{-1} \frac{a}{\lambda_d} , \quad  
\mu_d = \frac{\theta_0^2  \sigma \beta}{6 \eta \sqrt{D k}}, \quad   
 \lambda_d= \frac{6\lambda}{\theta_0} \,\sqrt{\frac{k}{D}}.
 \label{fricm} \end{equation}
In the limit $a \gg 1$ when the critical value $M_0$ in Eq.~(\ref{onset}) approaches the limit $M_0=4$, the droplet is mobile at $a < \lambda_d \exp (\mu_d/4)$. This suggests that the radius of mobile droplets is bounded both from below and above, as it indeed follows from the existence of a maximum in the radial dependence of the critical value of the parameter $\mu_d$, such as seen in Fig.~\ref{fmua}. 

Beyond the critical point, the velocity can be obtained using further terms in the expansion (\ref{woint0}). The dipole component of the term quadratic in $v$ vanishes, and therefore there is no contribution to motion in this order. The  dipole component of the third-order term is $\alpha_3 v^3$ where 
 \begin{equation}
\alpha_3=  - \frac{a^3}{16} \left[I_0(a)K_1(a)-I_1(a)K_0(a) - a^{-1}I_1(a)K_1(a)\right].                
   \label{on3} \end{equation}
This coefficient is negative; hence, the bifurcation is supercritical and propagation is possible at $M>M_0$. For small deviations $M_2=M-M_0>0$, the velocity is $v=\sqrt{-M_2/\alpha_3}$.   
%At $M_0 >4$, droplets exceeding a critical size (decreasing with increasing $M$) are mobile. 
%Droplets with very large radii may become immobile again due to the logarithmic growth of $M$ with increasing droplet size.   
%
% %%%%%%%%%%%%%%%%%%%%%%%%%%%%%%%%%%%%%%%%%%%%%%%%%
\subsection{Non-diffusive Limit \label{S34}}
% %%%%%%%%%%%%%%%%%%%%%%%%%%%%%%%%%%%%%%%%%%%%%%%%%

\begin{figure} [b]
\begin{center}
\begin{tabular}{cc}
(a) &  \\
& \includegraphics[width=8cm]{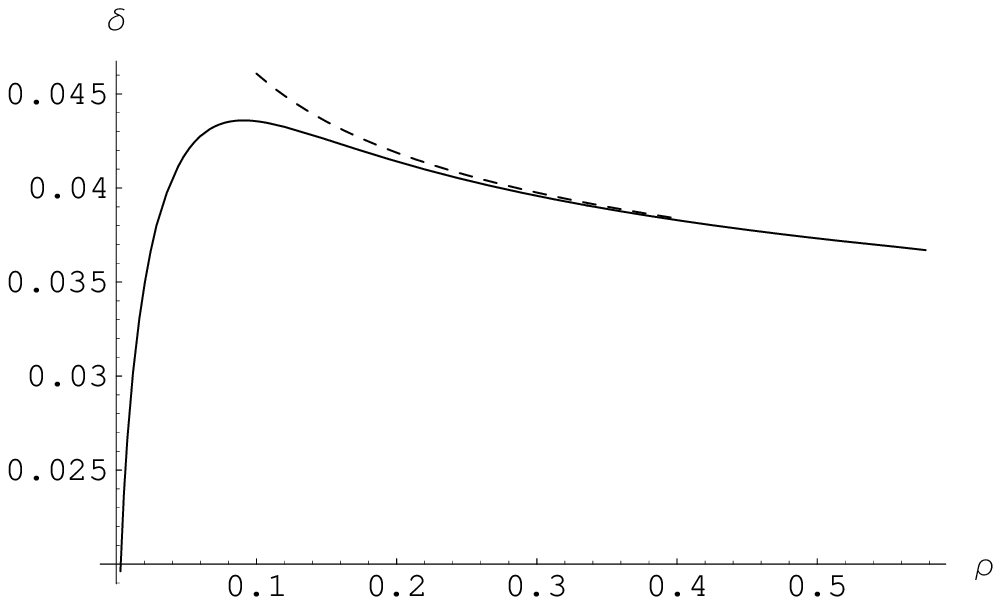}\\
(b) &  \\
& \includegraphics[width=8cm]{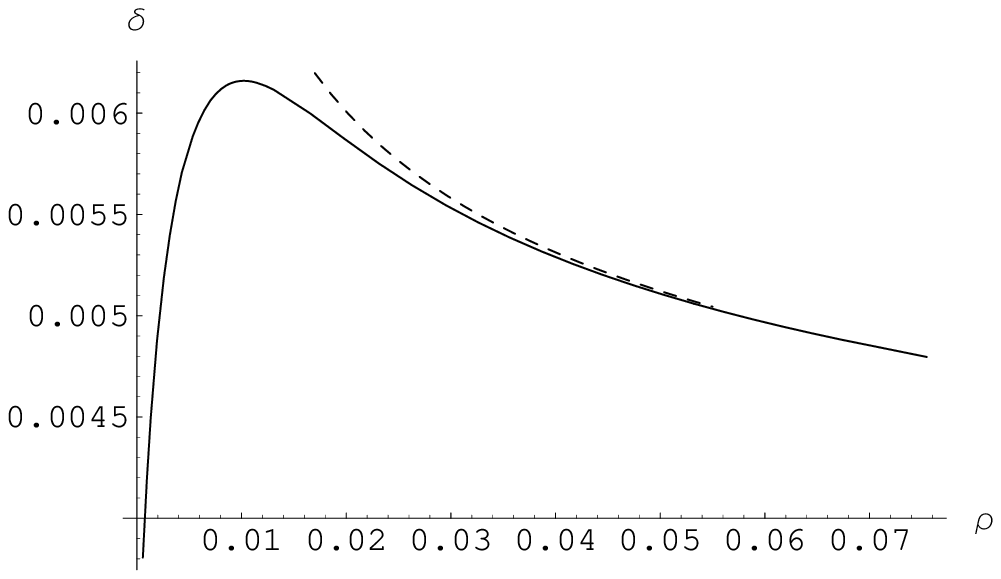} 
\end{tabular}
\end{center}
\caption{The dependence of the capillary number $\delta$ on the dimensionless radius $\rho$ for $\lambda_h=10^{-4}$ and $\mu_h=1$ (a) and $\mu_h=0.1$ (b). The dashed line shows the asymptotic dependence (\ref{vinf1}). 
\label{frv}}\end{figure}

In the opposite limit when diffusion is negligible, the surfactant concentration can be obtained directly by integrating along the direction of motion with the initial condition $c=1$ on the advancing contact line ($|\phi|<\pi/2$). The resulting concentration on the receding contact line,  
\begin{equation}
c (\phi) =  \E^{-2\tau \cos \phi},
 \label{cinf} \end{equation}
depends only on the ratio 
\begin{equation}
\tau = \frac{a}{v}= \frac{kR}{U} = \frac{\rho}{\delta}
 \label{tau} \end{equation}
where $\rho=Rk\eta/\gamma$ is the rescaled dimensionless droplet radius. The equation for $\tau$ following from Eq.~(\ref{fricd}) is
\begin{equation}
 P  =  \frac{ \tau}{\pi}  \int_0^{\pi/2} \cos \phi 
 \left(1- \E^{-2\tau \cos \phi} \right) \D \phi  ,
 \label{vinf} \end{equation}
containing a single parameter
\begin{eqnarray*}
P = \frac{a}{M}= \frac{6 \eta R k}{\theta_0^2  \sigma \beta} 
 \ln \frac{\theta_0 R}{6\lambda} 
 = \frac{\rho}{\mu_h}   \ln \frac{\rho}{\lambda_h}  , \\ \mbox{ where}\qquad
\mu_h = \frac{\beta\theta_0^2 }{6}, \qquad 
\lambda_h= \frac{6\lambda \gamma}{k\eta\theta_0}.
  \end{eqnarray*}
We rewrite Eq.~(\ref{vinf}) as
\begin{equation}
\frac{\rho}{\mu_h}  \ln \frac{\rho}{\lambda_h}  =  {\tau} F(\tau).
 \label{vinfq} \end{equation}
The function $F(\tau)$ is evaluated as
\begin{equation}
 F(\tau) =  \frac{1}{\pi} +  \frac{1}{2} 
 \left[I_1(2\tau) - \mathbf{L}_{-1}(2\tau) \right],
 \label{vinfst} \end{equation}
where $\mathbf{L}_{n}(x)$ is a Struve function. 
The function $F(\tau)$ increases monotonically from 0 at $\tau=0$ to $1/\pi$ at $\tau \to \infty$.

The dependence of velocity on the droplet radius can be obtained analytically in two limiting cases corresponding to the unsaturated and saturated regimes, respectively, at small and large $\tau$. In the former case, one can use the approximation $F(\tau) =  \tau/2 + O(\tau^2)$ to obtain
\begin{equation}
 \delta \approx   \left[ \frac{\rho \mu_h}{2\ln (\rho/\lambda_h)} \right]^{1/2}.
  \label{vinf0} \end{equation}
In the opposite limit $\tau \gg 1$, $F(\tau) \approx 1/\pi$ and
\begin{equation}
\frac{1}{\delta} \approx \frac{\pi}{\mu_h} \ln \frac{\rho}{\lambda_h} . 
  \label{vinf1} \end{equation}
Thus, the velocity increases with droplet size in the unsaturated and decreases in the saturated regime, in agreement with experiment \cite{e3} and earlier computations \cite{th05}. For intermediate values of $\tau$, the dependence of velocity on radius obtained by solving Eq.~(\ref{vinfq}) numerically is plotted in Fig.~\ref{frv}.

% %%%%%%%%%%%%%%%%%%%%%%%%%%%%%%%%%%%%%%%%%%%%%%%%%
\subsection{General case \label{S36}}
% %%%%%%%%%%%%%%%%%%%%%%%%%%%%%%%%%%%%%%%%%%%%%%%%%

\begin{figure} [h]
\begin{center}
\includegraphics[width=8.5cm]{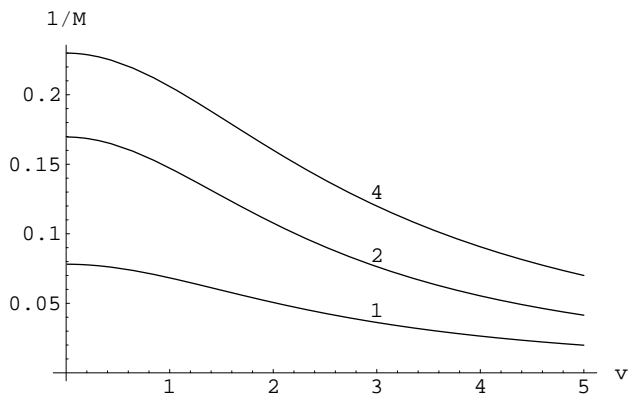} 
\end{center}
\caption{Plots of $G(a,v)$ defined by Eq.~(\ref{woinz}) as a function of $v$; the values of $a$ are marked at the respective curves. \label{fnum}}
\end{figure}

In a general case, Eq.~(\ref{fricv}) can be rewritten as $M^{-1} =G(a,v)$ and solved after computing numerically the double integral 
%
%\begin{widetext}
\begin{eqnarray}
&& G(a,v)  = \frac{a}{\pi^2 v}  \int_{0}^{\pi} \cos \phi \,\D \phi
 \int_{-\pi}^{\pi} 
   \E^{-\frac{1}{2} v a[\cos \phi- \cos (\phi+\zeta)]} \times \nonumber \\ 
 &&   \left[ \frac{ v}{2} \cos (\phi+\zeta) \,
 K_0\left(2a \sqrt{1+  \frac{v^2}{4}}
\sin \frac{|\zeta|}{2}  \right) \right. \nonumber \\ 
 && +\sqrt{1+  \frac{v^2}{4}}\, \sin  \frac{|\zeta|}{2}  
\left. K_1\left(2a \sqrt{1+  \frac{v^2}{4}}
\sin \frac{|\zeta|}{2}  \right)  \right] \D \zeta,
\label{woinz}  \end{eqnarray}
%\end{widetext}
%
The function $G(a,v)$ is plotted against $v$ at several values of $a$ in Fig.~\ref{fnum}. The curves peak at the ordinate at the bifurcation value $M_0^{-1}$ given by Eq.~(\ref{onset}).

% %%%%%%%%%%%%%%%%%%%%%%%%%%%%%%%%%%%%%%%%%%%%%%%%%
\section{Droplet Interactions  \label{S4}}
% %%%%%%%%%%%%%%%%%%%%%%%%%%%%%%%%%%%%%%%%%%%%%%%%%
%

% %%%%%%%%%%%%%%%%%%%%%%%%%%%%%%%%%%%%%%%%%%%%%%%%%
\subsection{Relaxation to a Stationary Pattern \label{S42}}
% %%%%%%%%%%%%%%%%%%%%%%%%%%%%%%%%%%%%%%%%%%%%%%%%%

\begin{figure} [b]
\begin{center}
\includegraphics[width=8.5cm]{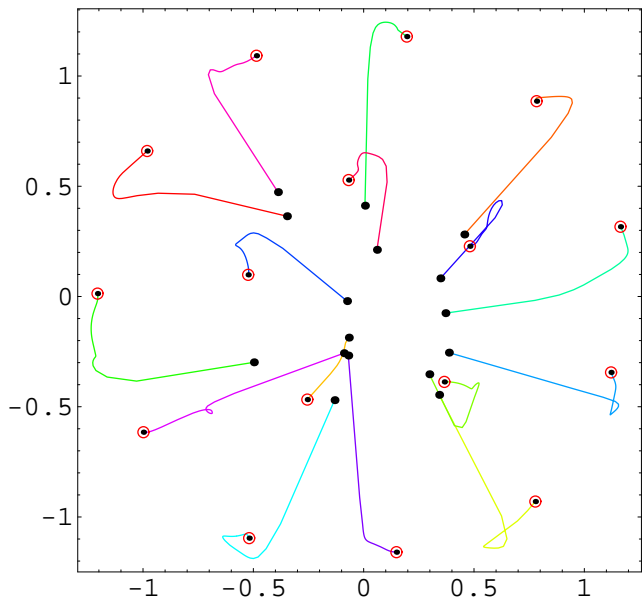} 
\end{center}
\caption{Trajectories of droplets moving according to  Eq.~(\ref{motion}) with added centripetal force. The dots mark the original random arrangement, and the circles, final positions. \label{freg}}
\end{figure}
At high diffusivities when droplets do not travel spontaneously, they still can move under the influence of mutual interactions. The surfactant depletion in the far field of a stationary droplet (at distances far exceeding its radius) is well approximated by the stationary solution Eq.~(\ref{diff}) with $c$ replaced by $\widetilde{c}$ and $H(\vect{x})$, by the delta-function multiplied by the droplet area: 
\begin{equation}
\widetilde{c} = \frac{1}{2} \,a^2\, K_0 (r).
  \label{cinf0} \end{equation}
The circular symmetry of the surfactant distribution around a single droplet is perturbed by the far field of its neighbors. The resulting repelling interaction  induces, according to Eq.~(\ref{fricd}), motion with the velocity proportional to the concentration gradient at the droplet location. If there is a number of droplets, their action is additive. This leads to the equation of motion for droplet centers $\vect{X}_j$
\begin{equation}
\frac{\D \vect{X}_j}{\D t} = M_j a_j \sum_{k \neq j} 
\frac{\vect{X}_j -\vect{X}_k}{|\vect{X}_j -\vect{X}_k|} \, \frac{a_k^2}{2} \,
  K_1 \left(|\vect{X}_j -\vect{X}_k|\right).
  \label{motion} \end{equation}
This is a gradient dynamical system
\begin{equation}
\frac{\D \vect{X}_j}{\D t} =  - \frac{M_j}{a_j} \frac{\partial V}{\partial \vect{X}_j},  \label{motion1} \end{equation}
evolving to minimize the potential 
\begin{equation}
V =  \frac{1}{2}\sum_{k \neq j} a_j^2 a_k^2 K_0 \left(|\vect{X}_j -\vect{X}_k|\right).
  \label{motion2} \end{equation}
In a confined region, the potential is expected to be minimized by a regular hexagonal pattern with spacing dependent on the number density of droplets. This is demonstrated by an example of evolution shown in Fig.~\ref{freg}. The confinement is effected in this computation by a centripetal external potential. One can see that evolution starting from a random arrangement of droplets evolves to regular pattern where circles mark final positions falling on a hexagonal grid. This might be a practical way to arrange a regular dewetting pattern on a homogeneous substrate.

% %%%%%%%%%%%%%%%%%%%%%%%%%%%%%%%%%%%%%%%%%%%%%%%%%
\subsection{Scattering \label{S41}}
% %%%%%%%%%%%%%%%%%%%%%%%%%%%%%%%%%%%%%%%%%%%%%%%%%

\begin{figure} [tb]
\begin{center}
\includegraphics[width=8.5cm]{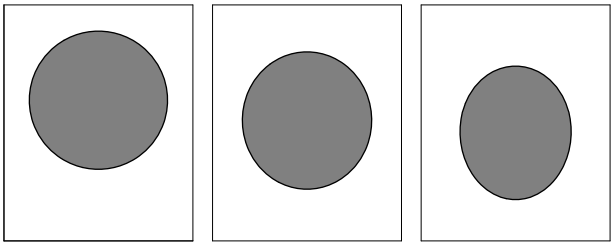} 
\end{center}
\caption{Surfactant depletion in the far field of a stationary droplet (left) and droplets propagating with the speed $v=1$ (center) and $v=2$ (right). The area with surfactant depletion above the same level is shaded, showing the depletion in the tail region increasing at higher speed. \label{fdist}}
\end{figure}

\begin{figure} [t]
\begin{center}
\includegraphics[width=8.5cm]{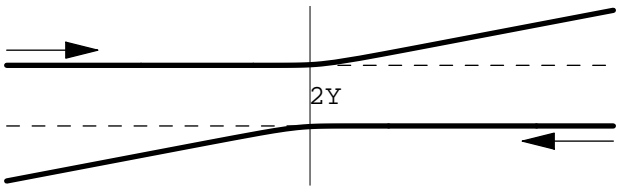} 
\end{center}
\caption{Scattering trajectories. The vertical line marks the location of the closest approach. \label{fsc}}
\end{figure}

\begin{figure} [b]
\begin{center}
 \includegraphics[width=8.5cm]{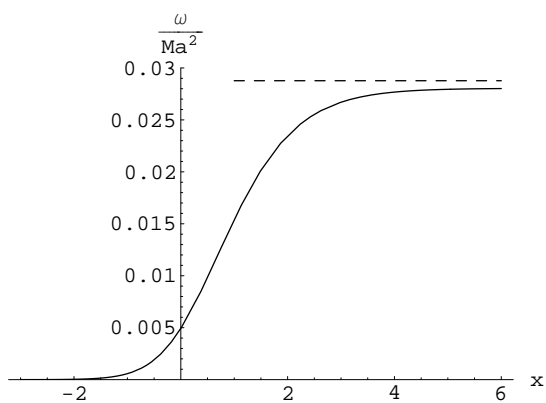}
 \end{center}
\caption{The change of the scattering angle with distance for $v=1,\; Y=2$. The dashed line shows the scattering angle at infinity computed with the help of Eq.~(\ref{sc2}). \label{fsc1}}
\end{figure}

\begin{figure} [t]
\begin{tabular}{c}
(a) \\ \includegraphics[width=8.5cm]{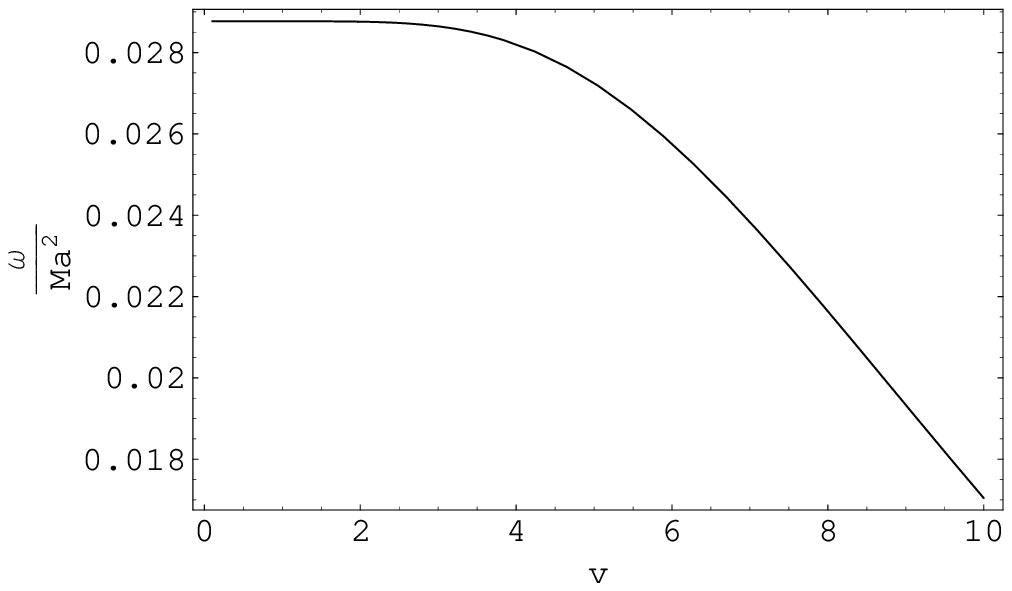} \\
(b) \\ \includegraphics[width=8.5cm]{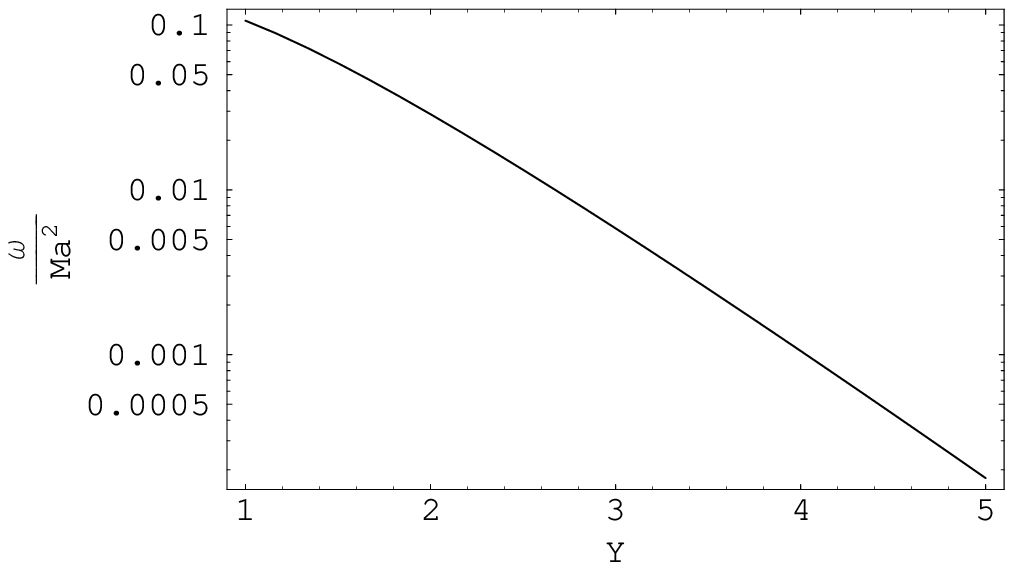} 
\end{tabular}
\caption{Dependence of the change of the scattering angle on velocity for $Y=0$  (a) and its dependence on the separation interval for $v=1$ (b). Both curves are computed using Eq.~(\ref{sc2}). \label{fscv}}
\end{figure}

The surfactant depletion in the far field of a steadily moving droplet can be obtained by solving  Eq.~(\ref{eqmot}) with $c$ replaced by $\widetilde{c}$ and $H(\vect{x})$, by the delta-function multiplied by the droplet area. The solution is expressed in polar coordinates $r,\phi$ centered on the droplet as
\begin{equation}
\widetilde{c} = \frac{a^2}{2}\, \E^{-\frac{1}{2} v r\cos \phi} 
  K_0\left(r \sqrt{1+  \frac{v^2}{4}} \right).
  \label{cinf} \end{equation}
The depletion field is strongly asymmetric, with a slower decay behind the droplet (Fig.~\ref{fdist}). 

Droplets moving one toward another are scattered by mutually repelling interaction created by the gradient of the far field. The problem remains tractable in the quasistationary approximation as long as velocity induced by interaction is much smaller than the speed of self-propelled motion. Otherwise, the far field becomes dependent on the entire history of motion, and solving the full non-stationary problem (\ref{diff}) is necessary.

Consider as an example two droplets of equal size propagating along the $x$ axis with identical speed $v$ on antiparallel trajectories shifted by the interval $2Y$, as in Fig.~\ref{fsc}. Since a deflected droplet keeps moving on a perturbed course, the scattering action is equivalent to acceleration in the direction normal to self-propelled motion.  Restricting to the quasistationary approximation, the dynamic equation for the deviation $\widetilde{y}$ normal to the original trajectory is therefore
\begin{equation}
\frac{\D^2 \widetilde{y}}{\D t^2} = - M a\,
 \frac{\partial \widetilde{c}}{\partial y} \,,
  \label{scm} \end{equation}
where the derivative of the surfactant depletion given by Eq.~(\ref{cinf}) is computed at a current distance between the droplets at the moment $t$ equal to
\begin{equation}
r(t) = 2\left[ (vt)^2 +\left(Y + \widetilde{y}\right)^2 \right]^{1/2}
  \label{sc1} \end{equation}
where the moment of closest approach is taken as $t=0$. Neglecting the change of the velocity component along the $x$-axis, time $t$ in Eq.~(\ref{scm}) can be replaced by $x/v$. A typical trajectory obtained by integration is shown in Fig.~\ref{fsc} and a more quantitative example of the change of the scattering angle with distance is shown in Fig.~\ref{fsc1}. Take note that, due to a faster decay of depletion ahead of the droplet, scattering largely accumulates already after the droplets have passed the point of closest approach.

For moderate deviations. a reasonable approximation for the scattering angle at infinity $\omega=\widetilde{y}'(\infty)$ gives the formula neglecting $\widetilde{y}$ compared to $Y$: 
\begin{widetext}
\begin{equation}
\omega = \frac{MYa^3}{2v}\, \sqrt{1+\frac{v^2}{4}} \int_{-\infty}^\infty 
\frac{ \E^{ v x} }{\sqrt{x^2 +Y^2}}
  K_1\left(r \sqrt{(4+ v^2)(x^2 +Y^2)} \right) \D x.
    \label{sc2} \end{equation}
\end{widetext}
Scattering angle computed with the help of this formula only weakly depends on velocity (Fig.~\ref{fscv}a). A much stronger dependence on the separation interval is shown in logarithmic coordinates in Fig.~\ref{fscv}b.

% %%%%%%%%%%%%%%%%%%%%%%%%%%%%%%%%%%%%%%%%%%%%%%%%%
\section{Surface Freezing and Melting  \label{S6}}
% %%%%%%%%%%%%%%%%%%%%%%%%%%%%%%%%%%%%%%%%%%%%%%%%%
%
A different mechanism of spontaneous motion has been observed in processes of surface freezing and melting \cite{r05}. In these experiments, liquid alkane wets silicon substrate better than a frozen smectic layer. Respectively, the equilibrium contact angle increases with growing thickness of smectic, and a droplet tends to slip to a lower level when placed at a terrace edge.

\begin{figure} [b]
\begin{center}
\begin{tabular}{cc}
(a) &  (b) \\
\includegraphics[width=4cm]{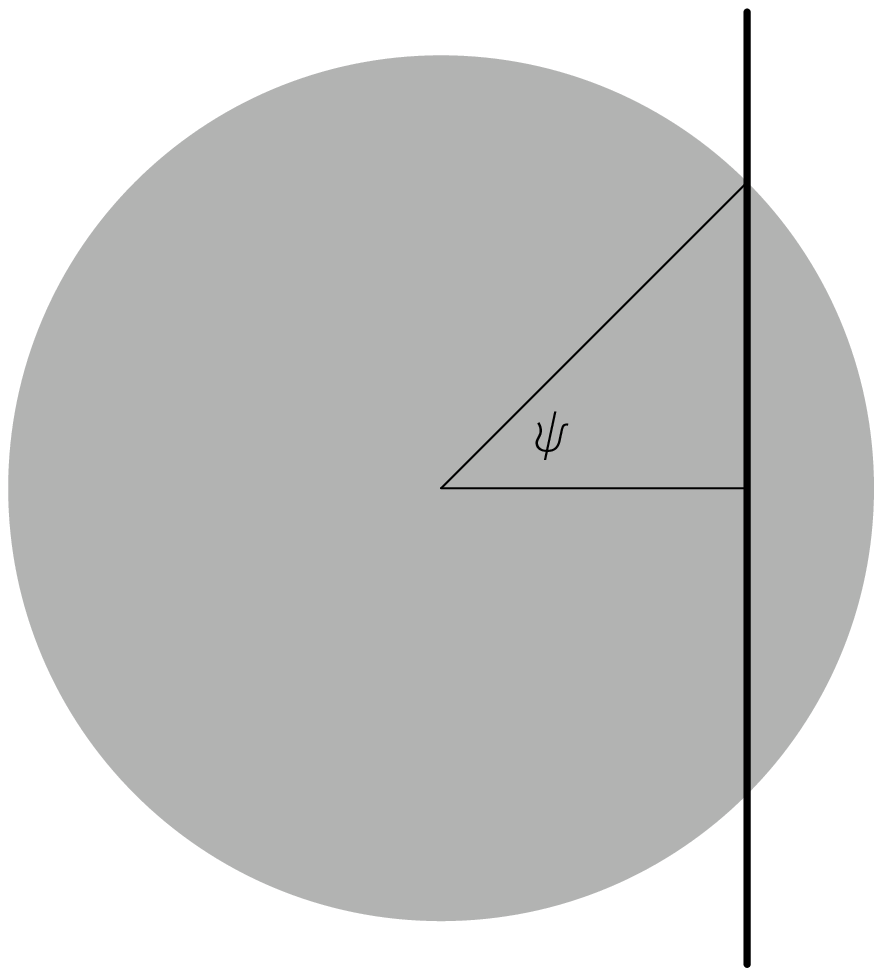} & 
\includegraphics[width=4.5cm]{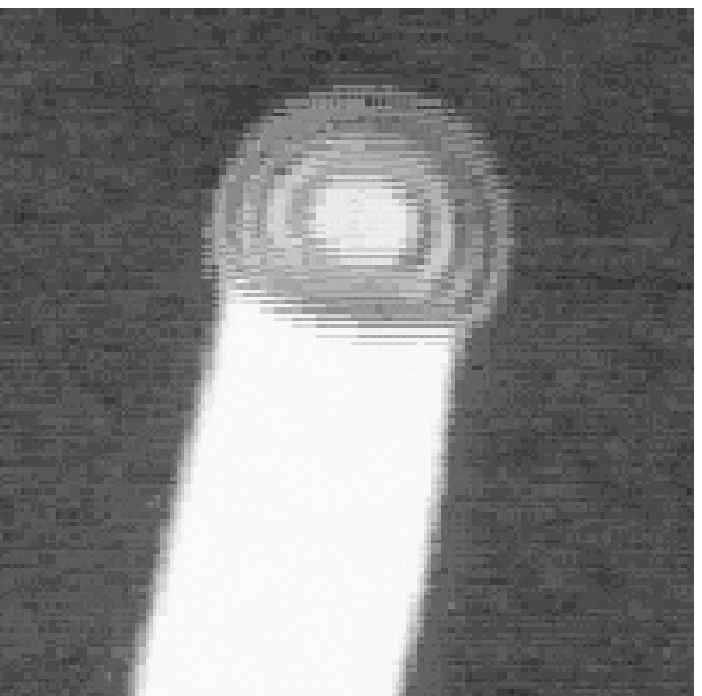} 
\end{tabular}
\end{center}
\caption{(a) A droplet on a terrace. (b) A snapshot from an experimental freezing sequence \cite{rc} showing a moving droplet leaving behind a frozen terrace. \label{fdrop}}
\end{figure}

For a droplet sitting on a terrace as shown in Fig.~\ref{fdrop}(a), $\widetilde \theta$ has distinct constant values on segments $|\phi|<\psi$ and $\psi<|\phi|<\pi$. The contact angle is smaller at the lower terrace, which we place on the right; we take this value as $\theta_0$ and denote the contact angle at the higher terrace as $\theta_0 + \widetilde \theta$. Then the velocity defined by Eq.~(\ref{fricd}) is 
\begin{equation}
U = U_0\sin \psi, \qquad 
U_0 = \frac{\gamma}{\eta}\, \frac{\theta_0^2}{6\pi} 
 \ln^{-1} \frac{\theta_0 R}{6\lambda} \widetilde \theta.
 \label{fricd1} \end{equation}
During freezing, the change of the angular front position $\psi$ due to propagation of the freezing front beneath the droplet with a constant velocity $C$ obeys $R \sin \psi \,\D \psi = - C \,\D t$. Due to the droplet motion, the net front velocity relative to the droplet center is $C-U$. Thus, the dynamic equation of the angular position is 
\begin{equation}
R\, \frac{\D \psi}{\D t} = U_0 - \frac{C}{\sin \psi}  .
 \label{fricd2} \end{equation}
The stationary position is $\psi_0 =\arcsin (C/U_0)$; $U_0$ is the maximum speed allowing for ths equilibrium. The configuration is stable at $\cos \psi <0$, i.e. $\psi >\pi/2$. Under this condition, the length of the line where the contact angles on the advancing and receding sides decreases when the droplet slides ahead, so that the freezing front catches up and the stationary configuration is restored. Another equilibrium position at $\psi<\pi/2$ is unstable.  This is in agreement with the disposition seen in Fig.~\ref{fdrop}(b) taken from the experimental freezing sequence \cite{rc}. The droplet travels forward, while its tail is hooked to the frozen terrace left behind. Stabilization at an obtuse angle, which is possible only in 3D, makes unnecessary a hypothetic synchronization mechanism through heat exchanged invoked in earlier 2D computations \cite{yp05}.    

One could expect such an equilibrium configuration to be impossible during melting when the directions of motion of the phase transition front and the droplet slip given by Eq.~(\ref{fricd}) are opposite. This equation, however, is not applicable during melting transition, since, unlike freezing when the back part of the droplet sits on the terrace it has created, the droplet formed as a result of melting is attached to the high-energy side surface of the smectic layer on the melting edge, and is carried along as this edge propagates \cite{r05}.     

%\paragraph*{Acknowledgement.} 
\acknowledgements
{This work has been supported by Israeli Science Foundation (grant 55/02). I thank Hans Riegler for discussions and access to his experimental data.}

%%%%%%%%%%%%%%%%%%%%%%%%%%%%%%%%%%%%%%%%%%%%%%%%%%%%

\end{document}